\documentstyle[12pt]{article}

\begin{document}
\baselineskip 0.7cm

\newcommand{\gsim}{ \mathop{}_{\textstyle \sim}^{\textstyle >} }
\newcommand{\lsim}{ \mathop{}_{\textstyle \sim}^{\textstyle <} }
\newcommand{\vev}[1]{ \left\langle {#1} \right\rangle }
\newcommand{\lsp}{ \left ( }
\newcommand{\rsp}{ \right ) }
\newcommand{\lmp}{ \left \{ }
\newcommand{\rmp}{ \right \} }
\newcommand{\llp}{ \left [ }
\newcommand{\rlp}{ \right ] }
\newcommand{\labs}{ \left | }
\newcommand{\rabs}{ \right | }
\newcommand{\KEV}{ {\rm keV} }
\newcommand{\MEV}{ {\rm MeV} }
\newcommand{\GEV}{ {\rm GeV} }
\newcommand{\TEV}{ {\rm TeV} }
\newcommand{\mgra}{m_{3/2}}
\newcommand{\tl}{\tilde{\lambda}}
\newcommand{\tv}{\tilde{v}}

\renewcommand{\thefootnote}{\fnsymbol{footnote}}
\setcounter{footnote}{1}

\begin{titlepage}

\begin{flushright}
TU-458
\\
May, 1994
\end{flushright}

\vskip 0.35cm
\begin{center}
{\Large \bf Flat Potential for Inflaton with a Discrete $R$-invariance in
Supergravity}
\vskip 1.2cm
Kazuya~Kumekawa, Takeo~Moroi
\footnote
{Fellow of the Japan Society for the Promotion of Science.}
and Tsutomu~Yanagida
\vskip 0.4cm

{\it Department of Physics, Tohoku University,\\
     Sendai 980-77, Japan}

\vskip 1.5cm

\abstract{
We show that a very flat potential of inflaton required for a
sufficient inflation is naturally obtained in $N=1$ supergravity by imposing a
discrete $R$-invariance $Z_{n}$. Several cosmological constraints on
parameters in the inflaton superpotential are derived. The reheating
temperature turns out to be $(1-10^8)$GeV for the cases of $n$=3--10.
Baryogenesis in this model is also discussed briefly.
}

\end{center}
\vspace{80 mm}
(Submitted to Prog. Theor. Phys.)
\end{titlepage}

\renewcommand{\thefootnote}{\arabic{footnote}}
\setcounter{footnote}{0}

%
%
%
%
\subsection*{1.~Introduction}

\hspace*{\parindent}
Much effort has been done in building realistic particle-physics models
based on $N=1$ supergravity~\cite{NPB212-413}. Although these theories
provide a natural framework for producing soft supersymmetry (SUSY)
breaking terms at low energies, there are potential cosmological
problems. One of them is a difficulty to construct inflationary
scenarios of the universe.  Namely, to generate a
sufficiently large expansion of the universe one must require an extreme
fine tuning of several parameters making a very flat potential for the
inflaton~\cite{sugra-inflation}. However, there has not been found any
natural explanation on the existence of such a flat potential.

In this paper, we show that a discrete $R$-symmetry $Z_{n}$
automatically leads to a flat potential at the origin of the inflaton
field $\phi$ as long as the minimum K\"ahler potential
$K(\phi,\phi^*)=\phi\phi^*$ is used.  We derive cosmological
constraints on parameters in the inflaton superpotential. 

In this model, the inflaton superpotential does not vanish at the
minimum of the inflaton scalar potential and hence the inflaton $\phi$
plays a role of the Polonyi field for generating a gravitino
mass~\cite{polonyi}.  Together with the cosmological constraints, all
relevant parameters in the superpotential are fixed such that the
gravitino mass $\mgra$ lies in 100GeV -- 10TeV.  With the obtained
parameters in the superpotential we argue that the most natural choice
for the discrete $R$-symmetry is $Z_4$, in which the reheating
temperature is $T_R\sim O(1-100)$TeV depending on the gravitino mass
$\mgra$=100GeV -- 10TeV. We give a brief comment on a possible
scenario for baryogenesis in such a low temperature universe.
Particle-physics problems related to the SUSY breaking in this model
are also discussed.

\subsection*{2.~Discrete $R$-symmetries and the inflaton potential}

\hspace*{\parindent}
Let us define a discrete $Z_{n}$ $R$-transformation on the inflaton field
$\phi (x,\theta)$ as
\begin{eqnarray}
\phi (x, \theta) \rightarrow
e^{-i \alpha} \phi (x, e^{i \alpha/2}\theta),
~~~\alpha = \frac{2\pi k}{n},
\label{r-sym}
\end{eqnarray}
with $k=0, \pm 1, \pm 2$$\cdot\cdot\cdot$.  Assuming positive power
expansion of a K\"ahler potential $K(\phi,\phi^*)$ and a superpotential
$W(\phi)$ for the inflaton field $\phi$, the general forms which have
the $Z_{n}$ $R$-invariance are given by
\begin{eqnarray}
K(\phi, \phi^{*}) 
= \sum_{m=1}^{\infty} a_{m}(\phi \phi^{*})^{m},
\label{kahler}
\end{eqnarray}
and 
\begin{eqnarray}
W(\phi) = \phi \sum_{l=0}^{\infty} b_{l} \phi^{ln}.
\label{super}
\end{eqnarray}
We take the minimum K\"ahler potential ($a_{1}$=1, other $a_{i}$=0) for
simplicity. (We have found that the scenario described below
does work with more general K\"ahler potential in eq.(\ref{kahler}), as
far as the coefficients $|a_2|<3/8n(n+1)$ and $|a_{i}|\sim O(1)
(i\geq3)$.)

In the $N=1$ supergravity~\cite{NPB212-413}, 
a scalar potential $V(\phi)$ is written as
\begin{eqnarray}
V(\phi) = \exp \lsp \frac{K(\phi, \phi^{*})}{M^{2}} \rsp
\lmp
(K^{-1})^{\phi^{*}}_{\phi} (D_{\phi} W) (D_{\phi} W)^{*}
- \frac{3 |W|^{2}}{M^{2}} \rmp, 
\label{pot}
\end{eqnarray}
with
\begin{eqnarray}
D_{\phi} W &\equiv&
\frac{\partial W}{\partial \phi} 
+ \frac{\partial K}{\partial \phi} \frac{W}{M^2},
\label{fterm} \\
(K^{-1})^{\phi^{*}}_{\phi} &\equiv&
\lsp \frac{\partial^{2} K}{\partial \phi \partial \phi^{*}} \rsp^{-1},
\end{eqnarray}
where $M$ is the gravitational mass $M=M_{Planck}/\sqrt{8\pi}\simeq
2.4\times 10^{18}\GEV$. With the general form of $W(\phi)$ in
eq.(\ref{super}), we find the inflaton potential $V(\phi)$ has
automatically  flat directions at the origin. That is, the condition
\begin{eqnarray}
\frac{\partial V}{\partial \phi} =
\frac{\partial^{2} V}{\partial \phi \partial \phi^{*}} =
0, 
\label{flat}
\end{eqnarray}
is always satisfied at the origin $\phi =0$. Notice that the mass term
$\phi\phi^*$ from the $\exp(K/M^2)$ is cancelled out by that
in the curly bracket in eq.(\ref{pot}). This cancellation is very
important for the inflaton $\phi$ to do its required job, and in all
previous works~\cite{sugra-inflation} this cancellation is
achieved only by a fine tuning of parameters in the superpotential
$W(\phi)$.\footnote
{There has been proposed an alternative approach, in which non-minimum
K\"ahler potentials are used~\cite{sugra-inflation2}. In this class of
models, a fine tuning is necessary in the K\"ahler potential to achieve
a very flat potential for the inflation.}

To find the approximate form of $V(\phi)$ near the origin
$\phi$$\sim$$0$ we write the superpotential as
\begin{eqnarray}
W(\phi) &=&
\lsp \frac{\lambda}{v^{n-2}} \rsp
\lsp v^{n} \phi - \frac{1}{n+1} \phi^{n+1} \rsp 
+ \cdot \cdot \cdot.
\label{super_n}
\end{eqnarray}
Here, $\cdot \cdot \cdot$ represents higher power parts
($\phi^{kn+1}$ with $k\geq 2$). Since these parts are all irrelevant
to the present analysis, we neglect them, hereafter. We easily see that
the inflaton potential (\ref{pot}) near $\phi\sim 0$ is well
approximated by
\begin{eqnarray}
V(\phi) \simeq \lsp \frac{\lambda}{v^{n-2}} \rsp^{2}
\lmp v^{2n} +\frac{1}{2} v^{2n} \lsp\frac{|\phi|^2}{M^2}\rsp^2 
-v^{n} (\phi^n + \phi^{*n}) \rmp.
\label{pot-ap}
\end{eqnarray}
Clearly, $n$=2 does not give a flat potential and hence we do not
consider the $n$=2 case. For $n$=3 and 4, we can neglect the
$(|\phi|^2/M^2)^2$ term, since $v\ll M$ as we will see later. For
$n\geq 5$, we also neglect this term assuming that the initial
amplitude of $\phi$ is greater than $v(v/M)^{4/(n-4)}$.\footnote
{This assumption on the initial amplitude $\varphi_{i}(\equiv\sqrt{2}{\rm
Re}\phi_i)$ is consistent with $\varphi_{i}<\varphi_{N}$(with $N\sim40$)
given in eq.(\ref{phin}) as far as $v\ll M$ and $n\ge5$. This 
initial condition $\varphi_{i}<\varphi_{N\sim40}$ is derived to have a
sufficient inflation as we will explain later.}
Thus, in any case of $n\geq 3$ we neglect the $(|\phi|^2/M^2)^2$ term in
analyzing the inflaton dynamics. In Fig.~1 we depict the
exact $V(\phi)$ along the real axis $\mbox{Re}\phi$, which shows that
the inflaton potential has always a
very flat region near $\phi\sim0$. We have also checked that $\phi\simeq
v$ is the global minimum of the inflaton potential.

Whether the inflation occurs or not depends on initial conditions for
the inflaton field $\phi$. In the present model the initial amplitude
of $\phi$ must be localized very near the origin in order to have a
sufficiently large expansion of the universe.  We simply assume
$\phi$ and $\dot{\phi}$ satisfy the desired initial conditions
$|\phi|\sim0$ and $\dot{\phi}\sim 0$.  We have no answer to a
question of what physics the initial conditions for $\phi$ was set
by, but if $\phi$ and $\dot{\phi}$ satisfy the initial condition,
the maximum inflation takes place and this exponentially expanding part
of the universe dominates the others.

Setting the phase of $\phi$ to vanish,\footnote
{Notice that the phase $\chi (x)~(\phi=\sqrt{2}\varphi\mbox{e}^{i\chi})$
has a positive mass at $\varphi\neq 0$. Thus if one chooses the
initial value of $\chi(x)\sim 0$, it stays there during the inflation.}
we identify the inflaton field with $\varphi$ where $\varphi$ is a real
component of $\phi$ ($\varphi\equiv{\sqrt 2}\mbox {Re}\phi$).
Thus, the relevant potential for $\varphi$ is now given by
\begin{eqnarray}
V(\varphi) \simeq \tl^{2} \tv^{4}
\lmp 1 - 2 \lsp \frac{\varphi}{\tv} \rsp^{n} \rmp,
\label{pot-phi}
\end{eqnarray}
with
\begin{eqnarray}
\tl \equiv\frac{1}{2}\lambda,~~~\tv \equiv\sqrt{2}v.
\end{eqnarray}

\subsection*{3.~Cosmological constraints}

\hspace*{\parindent}
The equation of motion for $\varphi$ in the expanding universe is given by
\begin{eqnarray}
\ddot{\varphi} + 3H\dot{\varphi} + \frac{dV}{d\varphi} = 0,
\label{eq-of-motion}
\end{eqnarray}
where $H$ is the Hubble expansion rate. During the slow rolling regime
of $\varphi$ ($\ddot{\varphi} \ll 3H\dot{\varphi}$) the energy density of the
universe is dominated by the inflaton potential $V(\varphi\sim 0)$, which
gives a nearly constant expansion rate
\begin{eqnarray}
H^2 \simeq \frac{V(\varphi\sim 0)}{3M^2}
\simeq \frac{\tl^2 \tv^4}{3M^2}.
\label{hubble}
\end{eqnarray}
In this inflationary epoch, the $\ddot{\varphi}$ term in
eq.(\ref{eq-of-motion}) can be neglected so that
\begin{eqnarray}
\dot{\varphi}
\simeq -\frac{V^{'}(\varphi)}{3H}
\simeq \frac{2 n \tl M}{\sqrt{3}\tv^{n-2}} \varphi^{n-1}.
\end{eqnarray}
The slow rolling regime ends at $\varphi_{f}$,
\begin{eqnarray}
\varphi_{f}^{n-2} \simeq \frac{3}{2n(n-1)} \frac{\tv^n}{M^2}.
\end{eqnarray}

The cosmic scale factor grows exponentially $\sim e^{N}$ till the end of
the inflation. With the above approximation, the $e$-folding factor $N$
between the time $t_{f}$ and $t_{N}$ is given by
\begin{eqnarray}
N=H(t_{f}-t_{N})\simeq
\frac{n-1}{3(n-2)}
\lmp \lsp \frac{\varphi_{N}}{\varphi_{f}}\rsp^{2-n} - 1 \rmp,
\end{eqnarray}
where $\varphi_{f}$ ($\varphi_{N}$) represents the amplitude of the field
variable $\varphi$ at the time $t_{f}$ ($t_{N}$). For a large $N$,
$\varphi_{N}$ is given by
\begin{eqnarray}
\varphi_{N} \simeq \lmp 2Nn(n-2) \rmp^{1/(2-n)}
\lsp \frac{\tv}{M} \rsp^{2/(n-2)} \tv.
\label{phin}
\end{eqnarray}
To solve the flatness and horizon problems, a sufficiently large
expansion of the universe is required during the
inflation~\cite{Kolb-Turner}. In our model, the Hubble radius of the
present universe crossed outside of the horizon $N\sim 40$ $e$-folds
before the end of the inflation.\footnote
{The reason why $e$-folding factor $N\sim 40$ is smaller than the usual
value $N\sim60$ is because the reheating temperature $T_R$ in the
present model is relatively low ($T_R\lsim 100\TEV$ for $n=3-5$). For
$n\geq6$, $T_R$ is $(10^2-10^4)\TEV$ and in these cases, $N$ becomes
$N\sim50$.}
This suggests that the
initial amplitude of $\varphi$ should be smaller than $\varphi_{N}$ with
$N\sim40$.\footnote
{This value of $\varphi_{N}$ is much larger than the quantum
fluctuation $\delta\varphi\sim H/2\pi$ unless $\lambda$ is large
$\lambda>(M/v)^\frac{n-4}{n-2}$.  Furthermore, the change of $\varphi$
in one expansion time is much larger than the quantum fluctuation.
Thus, the evolution of $\varphi$ can be discussed by solving the classical
equation of motion in eq.(\ref{eq-of-motion}).}

During the de~Sitter phase, the density perturbation
$(\delta\rho/\rho)$ arises from quantum fluctuations~\cite{fluc}
of the inflaton field $\varphi$. It is roughly given by
\begin{eqnarray}
\lsp \frac{\delta\rho}{\rho} \rsp_{N} \simeq
\frac{3}{5\pi} \frac{H^3}{\labs V^{'}(\varphi_{N})\rabs} \simeq
\lsp \frac{\tl \tv^3}{10\sqrt{3}\pi nM^3} \rsp
\lmp \frac{\tv^2}{2Nn(n-2)M^2} \rmp^{(1-n)/(n-2)}.
\end{eqnarray}
The relation between the density perturbation and the quadrupole of
the temperature fluctuation of cosmic microwave background (CMB)
$\sqrt{\langle a_2^2\rangle}$ is given by~\cite{APJ263-L1}
\begin{equation}
\sqrt{\langle a_2^2\rangle}=
\sqrt{\frac{5\pi}{12}}\lsp \frac{\delta\rho}{\rho} \rsp_{N\simeq40}.
\end{equation}
From the data on anisotropy of the CMB,
\begin{eqnarray}
\lsp\frac{\delta T}{T}\rsp\simeq
\sqrt{\frac{\langle a_2^2\rangle}{4\pi}}\simeq
6\times10^{-6}, 
\end{eqnarray}
observed by COBE~\cite{cobe}, we derive a constraint
\begin{eqnarray}
\left.
\frac{\tl \tv^3}{10\sqrt{3}\pi nM^3}
\lmp \frac{\tv^2}{2Nn(n-2)M^2} \rmp^{(1-n)/(n-2)}\rabs_{N\sim40}
 \simeq 2\times10^{-5}.
\label{dtpt}
\end{eqnarray}

We are now at the point to discuss the SUSY breaking in the present
model. Interesting is that the superpotential $W(\phi)$ does not
vanish at the potential minimum $\phi\simeq v$,
\begin{eqnarray}
W(\phi\simeq v) \simeq \frac{n}{\sqrt{2}(n+1)} \tl \tv^3.
\label{vev-of-W}
\end{eqnarray}
The gravitino mass $\mgra$ is, then, given by
\begin{eqnarray}
\mgra \simeq e^{\vev{K}/2M^2} \frac{\vev{W}}{M^2}
\simeq \frac{n}{\sqrt{2}(n+1)} \tl \tv \lsp \frac{\tv}{M} \rsp^2.
\label{mgra}
\end{eqnarray}
Thus, the inflaton field $\phi$ is regarded as the Polonyi field for
producing the gravitino mass~\cite{polonyi}.

From eqs.(\ref{dtpt}) and (\ref{mgra}), we determine $\lambda$ and $v$
for a given sets of $N$ and $\mgra$. The results for $N=40$ are shown in
Table~\ref{table1}.  For the obtained $\lambda$ and $v$, we calculate
the mass of $\phi$ as
\begin{eqnarray}
m_{\phi} \simeq \lambda n v.
\end{eqnarray}
As seen in Table~\ref{table1} the inflaton mass $m_{\phi}$ is
predicted as,
\begin{eqnarray}
m_{\phi} \simeq \lsp 10^{7} - 10^{12} \rsp \GEV.
\label{mphi-num}
\end{eqnarray}

This seems to contradict with the claim~\cite{PRD49-779} that the mass
of the Polonyi field is always at the gravitino mass scale
$\mgra\sim$100GeV -- 10TeV. However, our result (\ref{mphi-num}) is
not inconsistent with their claim, since we have not demanded the cosmological constant
to vanish. In fact, we have a non-zero cosmological constant at the
inflaton potential minimum,
\begin{eqnarray}
\Lambda^{\phi}_{\rm cos} = V(\phi \simeq v) \simeq -3 \mgra^2 M^2 
\sim -(10^{10} - 10^{11} \GEV)^4.
\label{lambda_cos}
\end{eqnarray}
Therefore, we need to invoke some mechanism to cancel this negative
cosmological constant. The simplest way is to introduce a U(1) gauge
multiplet $V(\theta, x)$ in the hidden sector and add a Fayet-Iliopoulos
$D$-term\footnote
{In supergravity, the Fayet-Iliopoulos $D$-term can be written
as (see Ref.\cite{wess-bagger} for notations)
\begin{eqnarray*}
\frac{3}{4} \int d^2 \Theta M^2 {\cal E} 
\lsp \bar{\cal D}\bar{\cal D} - 8R \rsp 
\exp \lsp - \frac{1}{3} \frac{\xi V}{M^2} \rsp .
\end{eqnarray*}
Clearly, this term is not invariant under the U(1) gauge transformation
$V \rightarrow V + \Lambda + \Lambda^{\dagger}$. Thus, we assume, here,
that this $D$-term is induced by some mechanism for the U(1) breaking.
Another solution to this problem may exist if the superpotential has a
continuous $R$-symmetry~\cite{PLB113-219,NPB211-302}. However, this
symmetry conflicts with our case, since the superpotential (\ref{super})
does not have the continuous $R$-symmetry. Therefore, we must also
consider that the continuous $R$-symmetry is broken down to our discrete
$R$-symmetry by some underlying physics at the Planck scale. A detailed
argument on this problem will be given in future
communication~\cite{moroi-yanagida}.}
~\cite{PLB51-461,PLB113-219,NPB211-302}, $\xi D$,
which shifts up the vacuum energy density by an amount of
\begin{eqnarray}
\delta\Lambda^{D}_{\rm cos} = \frac{1}{2} \xi^2.
\label{cctot}
\end{eqnarray}
Thus we can always choose $\xi$ so that the total cosmological constant
vanish,
\begin{eqnarray}
\Lambda_{\rm cos} 
= \Lambda^{\phi}_{\rm cos}+ \delta\Lambda^{D}_{\rm cos} =0.
\label{cc0}
\end{eqnarray}
This looks very artificial, but notice that a serious cosmological
problem~\cite{PLB131-59} associated with the light Polonyi field is not
present due to the relatively large mass $m_{\phi}$ given in
eq.(\ref{mphi-num}). In any case, the presence of non-vanishing
cosmological constant $\Lambda^{\phi}_{\rm cos}$ does not affect our
inflation scenario, since $\Lambda^{\phi}_{\rm cos}$ in
eq.(\ref{lambda_cos}) is always negligible\footnote
{Notice that we have not required, in our analysis, the cosmological
constant to be negligibly small. The main reason why we have
$\Lambda_{cos}^{\phi}\ll V(\phi\sim 0)$ is that the
vacuum-expectation value $v$ is very small, $(v/M)\sim
10^{-3}-10^{-2}$.}
compared with the inflaton
energy density $V(\phi\sim0)$ in the inflationary epoch, as seen in
Fig.~1.

Notice that the SUSY breaking due to the above $D$-term dominates over
the $F$-term breaking by the inflaton ($F=D_{\phi}W$, see
eq.(\ref{fterm})). Thus, the physical gravitino field $\psi_{\mu}^{'}$
is composed mainly of the $\psi_{\mu}$ and $\lambda$ through the super
Higgs mechanism~\cite{NPB212-413} as
\begin{eqnarray}
\psi_{\mu}^{'} = \psi_{\mu} 
+ \frac{1}{6} e^{-\vev{K}/2M^2} M \frac{\vev{D}}{\vev{W}}
\sigma_{\mu} \bar{\lambda} .
\end{eqnarray}
with
\begin{eqnarray}
\vev{D} = \xi,
\end{eqnarray}
where $\psi_{\mu}$ and $\lambda$ are the gravitino and the U(1) gaugino
fields, respectively (in the two component Weyl representation). This
physical gravitino field has a mass term
\begin{eqnarray}
e^{\vev{K}/2M^2}
\frac{\vev{W}}{M^2} \psi_{\mu}^{'} \sigma^{\mu\nu} \psi_{\nu} ^{'}
+ h.c.
\label{mgra2}
\end{eqnarray}
In the Minkowski space-time this gravitino mass becomes a physical pole
mass of the gravitino propagator.\footnote
{Notice that the presence of the gravitino mass given in eq.(\ref{mgra})
does not mean the SUSY breaking in the anti-de Sitter space-time.}
Furthermore, if we use the minimum K\"ahler potential for quark and
lepton fields, the soft-SUSY breaking masses for squarks and sleptons
come also from $W$ in eq.(\ref{vev-of-W}).

Since the inflaton $\phi$ couples to the light observable sector
very weakly (with strength $\sim (\lambda v^2/M^2)$), the decay width
$\Gamma_{\phi}$ is very small as
\begin{eqnarray}
\Gamma_{\phi} \sim \lsp \frac{\lambda v^2}{M^2} \rsp^2 m_{\phi}.
\end{eqnarray}
For the $n\leq 5$ case, this leads to a reheating temperature
$T_{R}$~\cite{Kolb-Turner},
\begin{eqnarray}
T_{R} = O(10^{-2}-10) \GEV,
\end{eqnarray}
which is too low to create enough baryon-number asymmetry of the
universe. 
In the case of $n\geq 6$, the reheating temperature is $O(100)$GeV. 
But as we will see later, models with $n\geq 5$ have a physical
cut-off smaller than the Planck scale. Therefore, if one requires the
cut-off scale to be larger than the Planck scale, one should take $n$=3
or 4 and, hence a new interaction is necessary to get a sufficiently
high reheating temperature for baryogenesis.

To have a faster decay of $\phi$, we introduce a new singlet chiral
supermultiplet $N(x, \theta)$ with a half $Z_{n}$ charge of $\phi$,
that is
\begin{eqnarray}
N(x,\theta) \rightarrow e^{-i \alpha/2} N (x, e^{i \alpha/2}\theta).
\end{eqnarray}
Then, we have a new interaction term in K\"ahler potential
\begin{eqnarray}
K(\phi,\phi^*,N,N^*) = \frac{g}{M} \phi^* NN + h.c.~,
\end{eqnarray}
and $N$ can have a mass term
\begin{eqnarray}
W(N) =  \frac{m_{N}}{2} NN.
\end{eqnarray}
Provided $2m_{N}<m_{\phi}$, we have a much faster $\phi$ decay mode,
$\phi\rightarrow NN$, with decay rate
\begin{eqnarray}
\Gamma_{\phi\rightarrow NN}
= \frac{g^2}{8\pi} \frac{m_{\phi}^3}{M^2} 
\lsp 1 - \frac{4m_{N}^2}{m_{\phi}^2} \rsp^2.
\end{eqnarray}
With this decay width, we estimate the reheating temperature $T_{R} \sim
(1 - 10^5)\TEV$ for $n$=4--10 (see Table~\ref{table1}).\footnote
{In the recent article Fischler have derived a new constraint on the
reheating temperature $T_{R}\lsim(10^2 - 10^5)\GEV$~\cite{Fischler} to
solve the gravitino problem~\cite{gravitino}. If one adopts this
constraint, there are left consistent only the $n=$3, 4 and 5 cases.
However, it is not clear to us if this new constraint on $T_{R}$ is
relevant.}
Only for the case of $n$=3, the reheating temperature seems to be too
low ($T_{R}\sim (1-100)\GEV$) for the baryogenesis as discussed below.

It is a straightforward task to incorporate the Fukugita-Yanagida
mechanism~\cite{PLB174-45} for baryogenesis in the present model,
identifying the singlet $N$ with three families of the right-handed
neutrinos.\footnote
{Through the seesaw mechanism~\cite{seesaw}, the light neutrino mass
$m_{\nu}$ is given by $m_{\nu}\simeq m_D^2/m_{N}$ with $m_D$ being the
Dirac mass of neutrino. If all $m_{N}\sim 10^9\GEV$ the MSW
solution~\cite{MSW} to the solar $\nu$ problem suggests
$m_D\simeq0.1-0.01\GEV$.}
 Taking $Z_{n}$ charges for all quarks $Q$ and leptons $L$ to be the same
as that of $N$ and assuming Higgs multiplets $H$ and $\bar{H}$ have
zero-$Z_{n}$ charges,\footnote
{The invariant mass term $W=\mu H\bar{H}$ is forbidden by the
$Z_{n}$-symmetry.  If one considers the exact $Z_{n}$-symmetry one needs
a Yukawa interaction $W=h\phi H\bar{H}$ to produce the invariant mass.
To give a week-scale mass for $H$ and $\bar{H}$ one must choose $h$ very
small, $h\sim10^{-14}-10^{-12}$.}
we find that the $N$'s can have the standard
Yukawa couplings
\begin{eqnarray}
W_{\rm Yukawa} = g_{ij} L_{i} N_{j} H,
\end{eqnarray}
where $i$ and $j$ represent the family indices. The decay
$N\rightarrow LH$ produces a lepton-number asymmetry if $g_{ij}$ has a
CP violating phase. The produced lepton-number asymmetry can be
converted to the baryon-number asymmetry~\cite{PLB174-45} through the
anomalous electroweak processes~\cite{PLB155-36} at high temperature
$T\gsim O(100\GEV)$.  Therefore, the reheating temperature should be
higher than $O(100\GEV)$.\footnote
{If $2m_{N_1}<m_{\phi}<2m_{N_{2,3}}$, only the $N_1$ is responsible for
the baryogenesis. Suppose a hierarchy in the Yukawa coupling,
$|g_{33}|\gg{\rm other}|g_{ij}|$, one obtains the lepton-number
asymmetry through the $N\rightarrow LH$ and  $L^*H^*$ decay
processes as~\cite{PLB174-45}
\begin{eqnarray*}
\Delta L/s \sim 10^{-5} \frac{1}{8\pi} |g_{33}|^2 \delta \lsp
\frac{m_{N_{1}}}{m_{N_{3}}} \rsp,
\end{eqnarray*}
where $10^{-5}$ is a dilution factor due to the entropy production of
$N$ decays, and $\delta$ is the CP-violating phase of $g_{ij}$. This
lepton-number asymmetry is converted to the baryon-number asymmetry and
one gets~\cite{PRD42-3344}
\begin{eqnarray*}
\Delta B/s \simeq \frac{4n_f+4}{12n_f+13} \Delta L/s ,
\end{eqnarray*}
with $n_f$ being the number of families ($n_f =3$).

With this result, one may easily explain the observed baryon-number
asymmetry $\Delta B/s\simeq 10^{-10} - 10^{-11}$, taking $g_{33}=O(1)$
and $\delta(m_{N_{1}}/m_{N_{3}})\sim10^{-4}$. Notice that the $N_1$
decay processes are always out of equilibrium since $m_{N_1}\gg T_R$ and
hence we do not have  any additional constraint (so-called
out-of-equilibrium condition) on $m_{N_1}$.}
As seen in Table~\ref{table1} this condition is satisfied in models with
$n\geq 4$, while in the $n=3$ case the gravitino mass less than
$O(10\TEV)$ is unlikely.\footnote
{If one uses the Affleck-Dine mechanism~\cite{NPB249-361} for the
baryogenesis, the $n=3$ case is not ruled out.}

\subsection*{4.~Conclusion}
\hspace*{\parindent}
Some comments are in order. 

(i) Since the superpotential $W(\phi)$
contains a non-renormalizable term $\frac{\lambda}{v^{n-1}}
\frac{1}{n+1}\phi^{n+1}$, the Born unitarity is violated in the process
$\phi + \phi^{*} \rightarrow (n-1)\phi+(n-1)\phi^{*}$
above a very high-energy scale. Thus we need a physical cut-off scale
$\Lambda$. By using a simple power counting, we estimate the
cut-off scale to be
\begin{eqnarray}
\Lambda \sim \lambda^{1/(1-n)} v.
\end{eqnarray}
If one imposes $\Lambda\geq M$ or $M_{Planck}$, one finds that only the
$n$=3 and $n$=4 cases are consistent (see Table~\ref{table1}). 

(ii) The gaugino masses come from the non-minimal kinetic term of
gauge multiplet $W_{\alpha}$~\cite{wess-bagger},
\begin{eqnarray}
\frac{1}{8g^2}\int d^2\Theta {\cal E} f(\phi) W^{\alpha}W_{\alpha} + h.c. ,
\label{kin-term}
\end{eqnarray}
where $f$ is the kinetic function for the gauge multiplet. 
Eq.(\ref{kin-term}) induces the gaugino mass term
\begin{eqnarray}
\frac{1}{4} \vev{\frac{\partial f}{\partial\phi} ~D_{\phi}W}
\tilde{g}^{\alpha}\tilde{g}_{\alpha} + h.c.
\end{eqnarray}
Suppose $f=\kappa\phi/M$, we get a very small gaugino mass\footnote
{We thanks M.Yamaguchi for pointing out this problem.} 
\begin{eqnarray}
m_{\tilde{g}} \simeq \frac{\kappa\lambda nv^4}{4(n+1)M^3} \sim
\lmp
\begin{array} {c}
10^{-3}\mgra,~~{\rm for}~n=4, \\
10^{-2}\mgra,~~{\rm for}~n=3,
\end{array}
\right.
&
\label{gmass3}
\end{eqnarray}
with $\kappa = O(1)$. Therefore, to obtain enough large gaugino
masses\footnote
{For $n$=3 and $\mgra$=10TeV, the predicted masses for gauginos in
eq.(\ref{gmass3}) are not excluded by the present experiment~\cite{pdg}.}
we must take account of radiative corrections from some heavy
particles\footnote
{The gravitino loop may also generate gaugino masses as pointed out in
Ref.~\cite{PLB113-219}}
~\cite{NPB221-29,PRD49-1446}. This situation is very much similar to in
the scenario of dynamical SUSY breaking~\cite{NPB256-557}.  Whether
the radiative corrections give rise to sufficiently large masses for the
gauginos depends on detailed physics at the high-energy (perhaps the
Planck) scale.  Therefore, it may be safe to conclude that the gaugino
masses are much smaller than those of the squarks and sleptons unless a
huge number of heavy particles exists at the high-energy scale.

In conclusion, we have shown that a discrete $R$-symmetry $Z_n$
naturally produces a very flat potential for the inflation.
Particularly for the case of $n=4$, the effective cut-off scale has
turned out to be at the Planck scale. Therefore, it will be interesting
to related our discrete $R$-symmetry to some physics at the Planck scale. 
A possible candidate is the superstring theory, since it is well known
that various discrete (non-$R$ and $R$-)symmetries arise from
compactified manifolds~\cite{superstring} of $10$-dimensional
space-time. Further investigation on this intriguing possibility will be
given elsewhere.

\subsection*{Acknowledgments}

We thank M.~Yamaguchi for useful discussions.

\newpage

%
%
\newcommand{\Journal}[4]{{\sl #1} {\bf #2} {(#3)} {#4}}
\newcommand{\APJ}{\sl Ap. J.}
\newcommand{\APJL}{\sl Ap. J. Lett.}
\newcommand{\CJP}{\sl Can. J. Phys.}
\newcommand{\NC}{\sl Nuovo Cimento}
\newcommand{\NP}{\sl Nucl. Phys.}
\newcommand{\PL}{\sl Phys. Lett.}
\newcommand{\PR}{\sl Phys. Rev.}
\newcommand{\PRL}{\sl Phys. Rev. Lett.}
\newcommand{\PTP}{\sl Prog. Theor. Phys.}
\newcommand{\SJNP}{\sl Sov. J. Nucl. Phys.}
\newcommand{\ZP}{\sl Z. Phys.}

\newpage
\subsection*{Figure caption}

{\bf Fig.~1} \\
{Inflaton potentials $V(\phi)$ along the real axis
Re$(\phi)$ are shown for the cases 
(a) $n=3$, $v=1.9\times 10^{16}$GeV,
(b) $n=4$, $v=2.5\times 10^{15}$GeV,
(c) $n=5$, $v=9.0\times 10^{14}$GeV. Notice
that these sets of parameters are given in
Table~\protect\ref{table1} with $\mgra$=1TeV and the shape of the
normalized potentials in this figure are independent of the parameter
$\lambda$.}

%
%
\newcommand{\ten}[2]{ $#1 \times 10^{#2}$ }

\begin{table}
    \begin{tabular}{rccccc}
	\multicolumn{6}{l} {$\mgra$=100GeV} \\ \hline
	$n$ & $\lambda$ & $v$[GeV] & $\Lambda$[GeV] &
		 $m_{\phi}$[GeV] & $T_R$[GeV] \\ \hline
	3 & \ten{6.5}{-10} & \ten{1.1}{16} & \ten{1.6}{25} & 
		\ten{2.1}{7} & {3.8} \\
	4 & \ten{5.0}{-7} & \ten{1.1}{15} & \ten{1.6}{18} & 
		\ten{2.3}{9} & \ten{4.4}{3} \\
	5 & \ten{1.3}{-5} & \ten{3.8}{14} & \ten{1.6}{16} & 
		\ten{2.5}{10} & \ten{1.6}{5} \\
	6 & \ten{8.9}{-5} & \ten{2.0}{14} & \ten{2.0}{15} & 
		\ten{1.1}{11} & \ten{1.4}{6} \\
	7 & \ten{3.1}{-4} & \ten{1.3}{14} & \ten{6.6}{14} & 
		\ten{2.8}{11} & \ten{5.9}{6} \\
	8 & \ten{7.2}{-4} & \ten{9.7}{13} & \ten{3.2}{14} & 
		\ten{5.6}{11} & \ten{1.7}{7} \\
	9 & \ten{1.3}{-3} & \ten{7.9}{13} & \ten{2.0}{14} & 
		\ten{9.5}{11} & \ten{3.7}{7} \\
	10 & \ten{2.1}{-3} & \ten{6.7}{13} & \ten{1.5}{14} & 
		\ten{1.4}{12} & \ten{6.9}{7} \\ \hline \\
	\multicolumn{6}{l} {$\mgra$=1TeV} \\ \hline
	$n$ & $\lambda$ & $v$[GeV] & $\Lambda$[GeV] &
		 $m_{\phi}$[GeV] & $T_R$[GeV] \\ \hline
	3 & \ten{1.2}{-9} & \ten{1.9}{16} & \ten{1.6}{25} & 
		\ten{6.6}{7} & \ten{2.1}{1} \\
	4 & \ten{5.0}{-7} & \ten{2.5}{15} & \ten{3.5}{18} & 
		\ten{4.9}{9} & \ten{1.4}{4} \\
	5 & \ten{9.9}{-6} & \ten{9.0}{14} & \ten{4.2}{16} & 
		\ten{4.4}{10} & \ten{3.7}{5} \\
	6 & \ten{5.6}{-5} & \ten{5.0}{14} & \ten{5.7}{15} & 
		\ten{1.7}{11} & \ten{2.7}{6} \\
	7 & \ten{1.7}{-4} & \ten{3.4}{14} & \ten{1.9}{15} & 
		\ten{4.1}{11} & \ten{1.1}{7} \\
	8 & \ten{3.7}{-4} & \ten{2.6}{14} & \ten{9.7}{14} & 
		\ten{7.8}{11} & \ten{2.8}{7} \\
	9 & \ten{6.5}{-4} & \ten{2.2}{14} & \ten{6.2}{14} & 
		\ten{1.3}{12} & \ten{5.7}{7} \\
	10 & \ten{9.9}{-4} & \ten{1.9}{14} & \ten{4.4}{14} & 
		\ten{1.9}{12} & \ten{1.0}{8} \\ \hline \\
	\multicolumn{6}{l} {$\mgra$=10TeV} \\ \hline
	$n$ & $\lambda$ & $v$[GeV] & $\Lambda$[GeV] &
		 $m_{\phi}$[GeV] & $T_R$[GeV] \\ \hline
	3 & \ten{2.1}{-9} & \ten{3.4}{16} & \ten{1.6}{25} & 
		\ten{2.1}{8} & \ten{1.2}{2} \\
	4 & \ten{5.0}{-7} & \ten{5.3}{15} & \ten{7.5}{18} & 
		\ten{1.1}{10} & \ten{4.4}{4} \\
	5 & \ten{7.4}{-6} & \ten{2.1}{15} & \ten{1.1}{17} & 
		\ten{7.9}{10} & \ten{8.8}{5} \\
	6 & \ten{3.5}{-5} & \ten{1.2}{15} & \ten{1.6}{16} & 
		\ten{2.7}{11} & \ten{5.5}{6} \\
	7 & \ten{9.7}{-5} & \ten{8.9}{14} & \ten{5.6}{15} & 
		\ten{6.0}{11} & \ten{1.9}{7} \\
	8 & \ten{1.9}{-4} & \ten{7.0}{14} & \ten{2.9}{15} & 
		\ten{1.1}{12} & \ten{4.5}{7} \\
	9 & \ten{3.2}{-4} & \ten{5.9}{14} & \ten{1.9}{15} & 
		\ten{1.7}{12} & \ten{8.8}{7} \\
	10 & \ten{4.6}{-4} & \ten{5.2}{14} & \ten{1.4}{15} & 
		\ten{2.4}{12} & \ten{1.5}{8} \\ \hline
    \end{tabular}
    \caption{For gravitino masses $\mgra$=100GeV, 1 and 10TeV,
             coupling constants $\lambda$ and vacuum-expectation values 
             $v$ are calculated with $(\delta T /T)=6.0\times10^{-6}$
             fixed. We have taken $n$=3--10, and 
             physical cut-off scales $\Lambda$, masses of $\phi$ field 
             $m_{\phi}$ and the reheating temperatures $T_{R}$ are also 
             shown for each $n$ .\label{table1}}
\end{table}
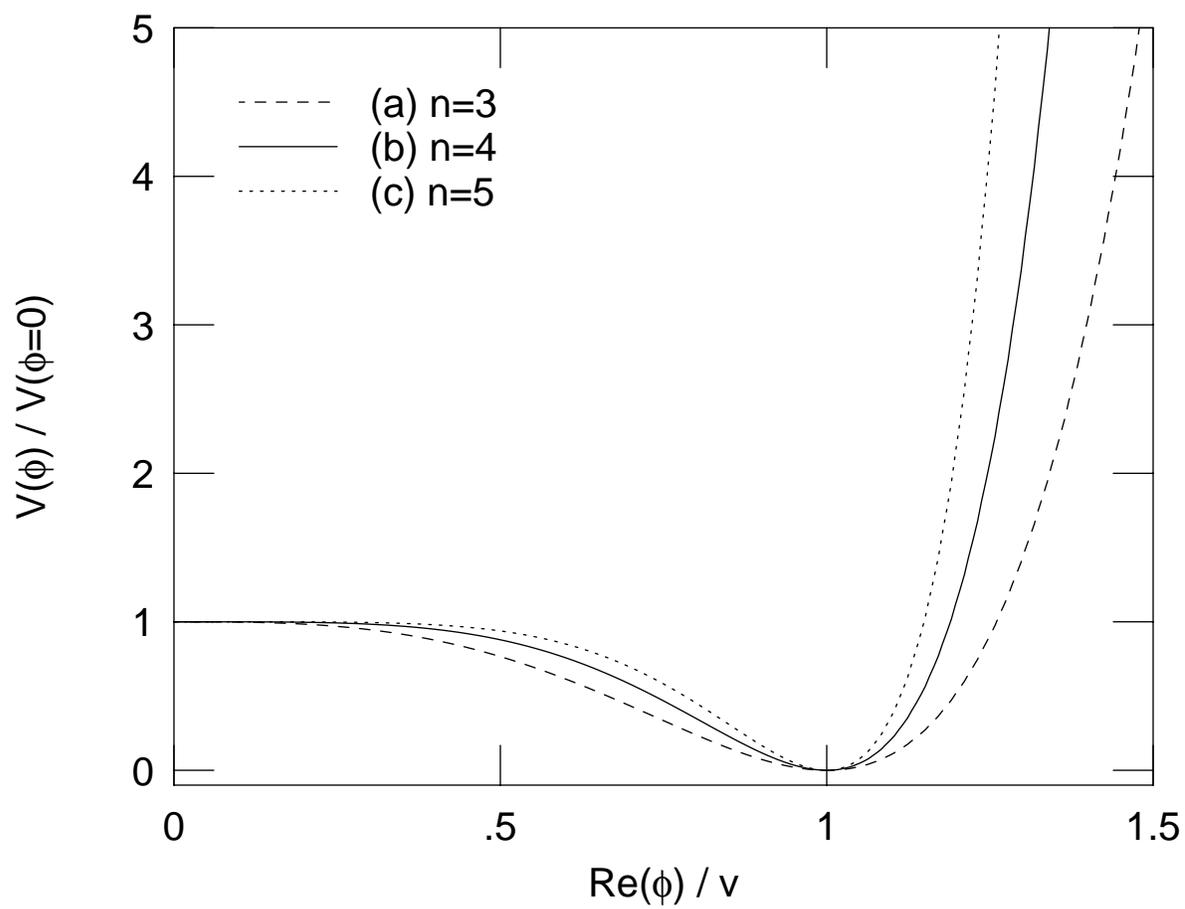
\end{document}